\documentclass[%
 reprint,
superscriptaddress,
showpacs,
 amsmath,amssymb,
aps,
prb,
]{revtex4-1}

\usepackage{graphicx}
\usepackage{dcolumn}
\usepackage{bm}

\usepackage{braket}
\usepackage{color}
\begin{document}

\preprint{APS/123-QED}

\title{The Holstein Polaron Problem Revisited}

\author{Amin Tayebi}
\email{tayebiam@msu.edu}
\affiliation{Department of Electrical and Computer Engineering, College of Engineering, Michigan State University, East Lansing, Michigan 48824, USA}
\affiliation{Department of Physics and Astronomy, Michigan State University, East Lansing, Michigan 48824, USA}
\author{Vladimir Zelevinsky}%
\affiliation{Department of Physics and Astronomy, Michigan State University, East Lansing, Michigan 48824, USA}
\affiliation{National Superconducting Cyclotron Laboratory, Michigan State University, East Lansing, Michigan 48824, USA}

\date{\today}

\begin{abstract}
The Holstein Hamiltonian was proposed half a century ago; since then, decades of research have come up empty handed in the pursuit of a closed-form solution. An exact solution to the two-site Holstein model is presented in this paper. The obtained results provide a clear image of the Hamiltonian structure and allow for the investigation of the symmetry, energy level crossings and polaronic characteristics of the system. The main mathematical tool is a three-term recurrence relation between the wave function amplitudes that was obtained using the properties of a family of orthogonal functions, namely the Poisson-Charlier polynomials. It is shown that, with the appropriate choice of basis, the eigenfunctions of the problem naturally fall into two families (parities) associated with the discrete $\mathbb{Z}_{2}$ symmetry of the Hamiltonian. The asymptotic solution to the recurrence relation is found by using the Birkhoff expansion. The asymptotic sets the truncation criterion for the wave function, which ensures the accurate calculation of the energy levels for any strength of electron-phonon interaction. Level crossing of states with different parities is discussed and the exact points of broken symmetry  are found analytically. The results are used as the building blocks for studying a four-site system. The inherited symmetries lead to the formation of a sparse matrix that is convenient for numerical calculations.
\end{abstract}

\pacs{63.10.+a, 63.20.-e, 63.20.kd}
\maketitle


\section{Introduction} \label{introduction}

The motion of charged carriers in polarizable media was first studied by Landau \cite{Landau}. The induced polarization together with the carrier form a quasiparticle, termed polaron \cite{Pekar}. Landau and Pekar depicted the picture as a stationary polaron trapped in its self-constructed potential well for sufficiently small phonon frequencies. They also noted an abrupt change for electrons of higher kinetic energy, when the polaron becomes mobile and propagates through the medium. It is now well understood that localization does not fully occur in a perfect lattice. Later, Fr\"{o}hlich {\sl et al.} developed a full quantum-mechanical description of so-called ``large" polarons in continuous media\cite{Frohlich} where the polaron wave function is extended over many lattice sites. A solution to the Fr\"{o}hlich Hamiltonian was developed by Feynman \cite{Feynman} using his path integral formulation. Fr\"{o}hlich's picture, however, breaks down in crystals where lattice discreteness cannot be ignored. Holstein was among the first to investigate the dynamics of polarons in crystal lattices\cite{Holstein}. The so-called Holstein ``small" polarons emerge due to the interaction between electrons and lattice vibrational modes (phonons). Detailed discussions on the large and small polarons can be found in\cite{Mitra,Devreese,Alexandrov0}.

The Holstein model has been adopted and utilized in a vast range of problems. In \cite{Alexandrov} the BCS theory was extended to high-temperature superconductors owing to polaron pairing (bipolaron mechanism). The polaronic approach has been used in \cite{Makarov} for studying the band structure of strongly correlated systems and in \cite{Leijnse} for investigating thermoelectric properties of molecular junctions. Over the past decade, the model has gained significant attention due to advanced technology making molecular-scale experiments possible. One such experiment is charge transport in organic semiconductors where polaronic effects play an important role \cite{Podzorov}. It is known that the presence of phonons in the system causes small resonant peaks in the transmission coefficient which can enhance the overall transmission \cite{Sapmaz,Zhang}.

Until now, all attempts to find a closed-form solution for the Holstein model has been fruitless. However, the model has been extensively studied using various theoretical techniques. Perturbative approaches have been applied to both the weak and strong electron-phonon interaction limits \cite{Marsiglio} and have led to satisfactory results. The perturbative methods can be accompanied by canonical transformations such as Lang-Firsov (LF) or modified Lang-Firsov (MLF) transformations in order to eliminate the direct electron-phonon coupling \cite{LangFirsov}. These transformations result in energy renormalization and incorporation of the electron-phonon interaction into the electronic hopping integral. Using LF and MLF transformations, it is shown in \cite{AlexandrovNEW,Chatterjee0} that perturbative approaches are applicable to a wide range of interaction strengths. More recently, a field-theoretical method was exploited in order to further extend the coverage of the coupling range \cite{Berciu}.

While these methods are accurate in the weak and strong regimes, they are less successful in describing the transitional situation. In order to overcome this problem, various numerical techniques such as exact diagonalization \cite{Wellein}, variational \cite{Bonca} and quantum Monte Carlo \cite{Hohenadler} algorithms have been exploited. Even though numerical calculations have significantly increased our understanding of polaronic physics, there is still a need for a rigorous analytical solution. Not only is this important from the mathematical point of view, it furthermore increases our insight into the subtleties of the problem. In addition, it can serve as the testing ground for numerical approaches yet to be developed. Accordingly, an attempt was made to obtain an analytical solution to the simplest polaronic system, namely the two-site Holstein problem, and the results are presented in this paper. The solution provides accurate results for all coupling strengths. The results can be used to greatly simplify the computational challenges encountered beyond the two-site model. This is illustrated in the case of a four-site system.

The paper is organized as follows. The Hamiltonian of the two-level system is considered in Section \ref{secII}. The symmetry of the Hamiltonian is discussed and it is shown that the eigenfunctions are divided into two parity groups.  With the proper choice of basis, the Schr\"{o}dinger equation reduces to a three-term recurrence relation for the wave function amplitudes. This leads to the accurate calculation of the energy levels. Next, the level crossing of different parity states associated with symmetry breaking is discussed; the exact crossing points are found analytically. Section \ref{secIII} considers the four-site Holstein model. With the system broken into two-level subsystems, the previous results can be directly applied. The symmetries of these two-level parts lead to an extremely sparse matrix convenient for numerical calculations. Section \ref{secIV} provides the conclusion. The first appendix contains a detailed calculation of the overlap functions. The latter two appendices introduce the Poisson-Charlier polynomials (a class of orthogonal functions) and the confluent Heun differential equation and discuss their properties used throughout the paper.

\section{The two-site system} \label{secII}

The two-site Holstein Hamiltonian with a single electron is
\begin{align} \label{D1}
H=& t(c_{L}^{\dagger} c_{R}+c_{R}^{\dagger} c_{L}) + \omega(b_{L}^{\dagger} b_{L}+b_{R}^{\dagger} b_{R}) \nonumber \\
+& \sqrt{2}g \left(c_{L}^{\dagger} c_{L}(b_{L}^{\dagger}+b_{L})+c_{R}^{\dagger} c_{R}(b_{R}^{\dagger}+b_{R})\right),
\end{align}
where $c_{L/R}^{\dagger}$($b_{L/R}^{\dagger}$) and $c_{L/R}$($b_{L/R}$) are fermionic (bosonic) creation and annihilation operators for the Left/Right site, respectively. The parameter $t$ is the electron hopping amplitude between the sites, $\omega$ is the phonon frequency and $g$ is the on-site electron-phonon interaction strength. The factor of $\sqrt{2}$ is chosen for convenience. It is useful to introduce new phonon operators $b=(b_{L}-b_{R})/\sqrt{2}$ and $\mathcal{B}=(b_{L}+b_{R})/\sqrt{2}$. Then the transformed Hamiltonian can be separated into two commuting pieces. The first piece is a shifted oscillator, $\omega \mathcal{B}^{\dagger}\mathcal{B}+g(\mathcal{B}^{\dagger}+\mathcal{B})$, with energy eigenvalues $\omega n-\frac{g^{2}}{\omega}$ where $n$ is a non-negative integer. The second piece, revealing that at the core of the Holstein Hamiltonian lies a two-level fermionic system interacting with a bosonic field, is expressed as
\begin{equation} \label{D2}
H_{r}=t(c_{L}^{\dagger} c_{R}+c_{R}^{\dagger} c_{L}) + \omega b^{\dagger} b+g(c_{L}^{\dagger} c_{L}-c_{R}^{\dagger} c_{R})(b^{\dagger}+b).
\end{equation}

The analogy of the left and right sites to the upper and lower states of a qubit can be used in order to translate the Hamiltonian $H_{r}$ into the language of quantum electrodynamics. It is easy to see that the analogous translated Hamiltonian is
\begin{equation}
H'_{r}=t\sigma_{x} + \omega b^{\dagger} b+g\sigma_{z}(b^{\dagger}+b),
\end{equation}
where $\sigma_{x}$ and $\sigma_{z}$ are the Pauli $2\times 2$ matrices. Next, by applying a unitary transformation $\mathcal{U}=\frac{1}{\sqrt{2}}(\sigma_{x}+\sigma_{z})$ to $H'_{r}$, one comes to the popular quantum Rabi
Hamiltonian\cite{Rabi},
\begin{equation}
H_{R}=t\sigma_{z} + \omega b^{\dagger} b+g\sigma_{x}(b^{\dagger}+b).
\end{equation}
The connection between the two models allows us to borrow ideas previously developed for studying the Rabi Hamiltonian. An important gain is the $\mathbb{Z}_{2}$-symmetry of the Rabi model \cite{Schiro}. In our language, the parity operator
\begin{equation}
\mathcal{P}=e^{i\pi Q}, \quad {\rm where} \quad  Q=b^{\dagger}b+\frac{1}{2}(c_{L}^{\dagger} c_{R}+c_{R}^{\dagger} c_{L}+1),
\end{equation}
commutes with the Hamiltonian (\ref{D2}). To see this, note that
\begin{align} \label{D3}
\mathcal{P}b\mathcal{P}^{\dagger} &=-b , \nonumber \\
\mathcal{P}(c_{L}^{\dagger} c_{L}-c_{R}^{\dagger} c_{R})\mathcal{P}^{\dagger}&=-(c_{L}^{\dagger} c_{L}-c_{R}^{\dagger} c_{R}) , \nonumber \\
\mathcal{P}(c_{L}^{\dagger} c_{R}+c_{R}^{\dagger} c_{L})\mathcal{P}^{\dagger}&=+(c_{L}^{\dagger} c_{R}+c_{R}^{\dagger} c_{L}),
\end{align}
which illustrates that the Hamiltonian is invariant under $\mathcal{P}$, i.e. $\mathcal{P}H_{r} \mathcal{P}^{\dagger}=H_{r}$. Therefore the eigenfunctions of the Hamiltonian (\ref{D2}) can be classified by parity.
In order to obtain the eigenfunctions and corresponding energy levels, we must solve the Schr\"{o}dinger equation.

\subsection{Solving the Schr\"{o}dinger equation}

For the appropriate choice of a basis, we define two types of new phonon operators,
\begin{equation}
\mathcal{B}_{\pm}=b\pm\frac{g}{\omega},
\end{equation}
with eigenstates $\phi^{\pm}(n)$, where $n$ is a non-negative integer counting the number of corresponding
phonons. Therefore
\begin{eqnarray} \label{D4}
\Big[\omega b^{\dagger}b+g(b^{\dagger}+b)\Big]\phi^{+}(n)&&=\left(\omega \mathcal{B}^{\dagger}_{+}\mathcal{B}_{+}-\frac{g^{2}}{\omega}\right)\phi^{+}(n) \nonumber\\
&& =\left(\omega n-\frac{g^{2}}{\omega}\right)\phi^{+}(n),
\end{eqnarray}
and
\begin{eqnarray} \label{D5}
\Big[\omega b^{\dagger}b-g(b^{\dagger}+b)\Big]\phi^{-}(n)&&=\left(\omega \mathcal{B}^{\dagger}_{-}\mathcal{B}_{-}-\frac{g^{2}}{\omega}\right)\phi^{-}(n) \nonumber\\
&& =\left(\omega n-\frac{g^{2}}{\omega}\right)\phi^{-}(n).
\end{eqnarray}
Using $\phi^{+}(n)$ and $\phi^{-}(n)$ as the basis for phonons, a general ansatz for the stationary wave
fuinction of the Hamiltonian (\ref{D2}) is
\begin{equation}
\psi=c^{\dagger}_{L}\sum_{n=0}^{\infty}a_{L}(n)\phi^{+}(n)+c^{\dagger}_{R}\sum_{n=0}^{\infty}a_{R}(n)\phi^{-}(n).
\end{equation}
Now the Schr\"{o}dinger equation yields
\begin{eqnarray} \label{D6}
&& c^{\dagger}_{L} \sum_{n=0} \Big[ ta_{R}(n)\phi^{-}(n)+a_{L}(n) \Big( \omega n -\frac{g^{2}}{\omega} - \mathcal{E} \Big) \phi^{+}(n) \Big] \nonumber\\
+ && c^{\dagger}_{R} \sum_{n=0}  \Big[ ta_{L}(n)\phi^{+}(n)+a_{R}(n) \Big( \omega n -\frac{g^{2}}{\omega} - \mathcal{E} \Big) \phi^{-}(n) \Big]\nonumber\\
&&=0 ,
\end{eqnarray}
where $\mathcal{E}$ is the energy.
In order to obtain algebraic equations for the coefficients $a_{L}(n)$ and $a_{R}(n)$, Eq.~(\ref{D6}) is projected onto $\bra{\phi^{+}(m)}$ and $\bra{\phi^{-}(m)}$. This results in the first pair of equations,
\begin{align} \label{D7}
t\sum_{n=0}^{\infty}\Big[ a_{R}(n) P^{+-}_{mn}\Big] + a_{L}(m) \Big(\omega m -\frac{g^{2}}{\omega} - \mathcal{E}\Big) &=0 , \nonumber \\
ta_{L}(m)+\sum_{n=0}^{\infty}\Big[a_{R}(n)\Big(\omega n -\frac{g^{2}}{\omega} - \mathcal{E}\Big)P^{+-}_{mn}\Big] &=0,
\end{align}
and the second pair,
\begin{align} \label{D8}
t\sum_{n=0}^{\infty}\Big[a_{L}(n) P^{-+}_{mn}\Big] + a_{R}(m) \Big(\omega m -\frac{g^{2}}{\omega} - \mathcal{E}\Big) &=0, \nonumber \\
ta_{R}(m)+\sum_{n=0}^{\infty}\Big[a_{L}(n)\Big(\omega n -\frac{g^{2}}{\omega} - \mathcal{E}\Big)P^{-+}_{mn}\Big] &=0,
\end{align}
where the overlap functions are defined as
\begin{equation}
P^{+-}_{mn}=\bra{\phi^{+}(m)}\phi^{-}(n)\rangle, \quad P^{-+}_{mn}=\bra{\phi^{-}(m)}\phi^{+}(n)\rangle,
\end{equation}
(for their explicit expressions see appendix A). Decoupling Eqs.~(\ref{D7}) and (\ref{D8}) leads to the two uncoupled equations:
\begin{align} \label{D9}
& t^{2}a_{L}(m)=\nonumber\\
&\sum_{n,n'}a_{L}(n')\Big(\omega n' -\frac{g^{2}}{\omega} - \mathcal{E}\Big)\Big(\omega n -\frac{g^{2}}{\omega} - \mathcal{E}\Big)P^{-+}_{nn'}P^{+-}_{mn},
\end{align}
and
\begin{align} \label{DD9}
& t^{2}a_{R}(m)=\nonumber\\
&\sum_{n,n'}a_{R}(n')\Big(\omega n' -\frac{g^{2}}{\omega} - \mathcal{E}\Big)\Big(\omega n -\frac{g^{2}}{\omega} - \mathcal{E}\Big)P^{+-}_{nn'}P^{-+}_{mn}.
\end{align}

The overlap functions can be written in terms of the Poisson-Charlier polynomials, see appendix B. This provides the necessary tools for simplifying Eqs.~(\ref{D9}) and (\ref{DD9}) . Using the orthogonality (\ref{B2}) and recurrence (\ref{B3}) relations of the polynomials, Eq.~(\ref{D9}) reduces to
\begin{align} \label{D10}
\zeta_{m}^{1} a_{L}(m+1) - \zeta_{m}^{0} a_{L}(m) + \zeta_{m}^{-1} a_{L}(m-1)=0 ,
\end{align}
and Eq.~(\ref{DD9}) reduces to
\begin{align} \label{DD10}
\zeta_{m}^{1} a_{R}(m+1) + \zeta_{m}^{0} a_{R}(m) + \zeta_{m}^{-1} a_{R}(m-1)=0 ,
\end{align}
where the coefficients are
\begin{align}
\zeta_{m}^{1}&=2g\sqrt{m+1}\Big[\omega (m+1)-\frac{g^2}{\omega}-\mathcal{E}\Big], \nonumber \\
 \zeta_{m}^{0}&=\Big(\omega m-\frac{g^2}{\omega}-\mathcal{E}\Big)\Big(\omega m+\frac{3g^2}{\omega}-\mathcal{E}\Big)-t^{2}, \nonumber \\
\zeta_{m}^{-1}&=2g \sqrt{m}\Big[\omega (m-1)-\frac{g^2}{\omega}-\mathcal{E}\Big] .
\end{align}
The corresponding boundary conditions are $a_{L}(-1)=0$ and $a_{L}(0)$ an arbitrary number which we set equal to unity. The parity symmetry discussed earlier is noted in (\ref{D10}) and (\ref{DD10}): there exist only two types of eigenfunctions, the first type with $a_{L}(m)=(-)^{m}a_{R}(m)$ and the second type with $a_{L}(m)=(-)^{m+1}a_{R}(m)$.

It is useful to introduce a new variable $a_{L}(m)=\sqrt{m!}(\frac{w}{2g})^{m}D(m)$. The recurrence relation (\ref{D10}) in terms of $D(m)$ is
\begin{align} \label{D11}
 & (m+1)\Big(m+1-\frac{g^{2}}{\omega^{2}}-\frac{\mathcal{E}}{\omega}\Big) D(m+1) = \nonumber \\
& \Big[\Big(m-\frac{g^{2}}{\omega^{2}}-\frac{\mathcal{E}}{\omega}\Big) \Big(m+\frac{3g^{2}}{\omega^{2}}-\frac{\mathcal{E}}{\omega}\Big)-\frac{t^{2}}{\omega^{2}} \Big]D(m) \nonumber \\
& -\Big(\frac{2g}{\omega}\Big)^{2}\Big(m-1-\frac{g^{2}}{\omega^{2}}-\frac{\mathcal{E}}{\omega}\Big)D(m-1),
\end{align}
with the boundary conditions $D(-1)=0$ and $D(0)=1$.

The recurrence relation (\ref{D11}) is of  Poincar\'{e} type and allows for the use of the Birkhoff asymptotic expansion \cite{Wimp}. Following \cite{Wong} the leading order asymptotic for $D(m)$ is
\begin{equation} \label{D12}
D(m)\sim \frac{c_{0}}{m!} \left(\frac{2g}{\omega}\right)^{2m} \ \frac{m^{\frac{g^{2}}{\omega^{2}}+\frac{\mathcal{E}}{\omega}}}{m-\frac{g^{2}}{\omega^2}-\frac{\mathcal{E}}{\omega}}, \ m\rightarrow \infty ,
\end{equation}
for some constant $c_{0}>0$. In practical calculations, one can evaluate (\ref{D11}) recursively. The asymptotic should be used as a cut-off criterion. It is important to truncate the series only after the asymptotic has reached its maximum value. An exact solution to the Holstein two-site problem in the form of recursive relations was previously obtained in \cite{Rongsheng}. However, their recurrence relation involved an infinite series of the wave function amplitudes which, compared to Eq. (\ref{D11}), makes it less convenient for practical calculations.

\subsection{Energy levels and polaronic characteristics}

In order to obtain the energy levels, we go back to Eq.~(\ref{D7}). We multiply both sides of the two equations by $\left(\frac{g}{\omega}\right)^{m}/\sqrt{m!}$ and then sum over $m$ from $0$ to $\infty$. Using the generating function identity (\ref{B4}) we have an equation for eigenvalues $\mathcal{E}$ 
\begin{equation} \label{D13}
\mathcal{K}^{\pm}(\mathcal{E})=\sum_{m=0}^{\infty}2^{-m}D(m)\Big( \omega m-\frac{g^{2}}{\omega}-\mathcal{E}\pm t \Big)=0 ,
\end{equation}
where positive and negative signs correspond to the two types of states with $a_{L}(m)=(-)^{m}a_{R}(m)$ and $a_{L}(m)=(-)^{m+1}a_{R}(m)$, respectively. The three-term recurrence relation (\ref{D11}) together with condition (\ref{D13}), provide the energy levels of the Hamiltonian (\ref{D2}). A straightforward method to calculate the energies is to first evaluate  $\mathcal{K}^{\pm}$ as a function of a continuous real parameter $x$ over the range of interest. The energy levels are then the zero points of $\mathcal{K}^{\pm}(x)$. Fig.~1 shows $\mathcal{K}^{+}(x)$ (black) and $\mathcal{K}^{-}(x)$ (red or gray) for $g=0.6$ and the adiabatic parameter $t/ \omega=0.7$. The intersection points, where $\mathcal{K}^{\pm}(x)=0$, are the energy levels of the system. The black and red (gray) curves in Fig.~1 correspond to states with positive and negative parity, respectively. It was noted in \cite{Schiro} that the ground state has a fixed positive parity.
\begin{figure}[h]
\includegraphics[width=8cm]{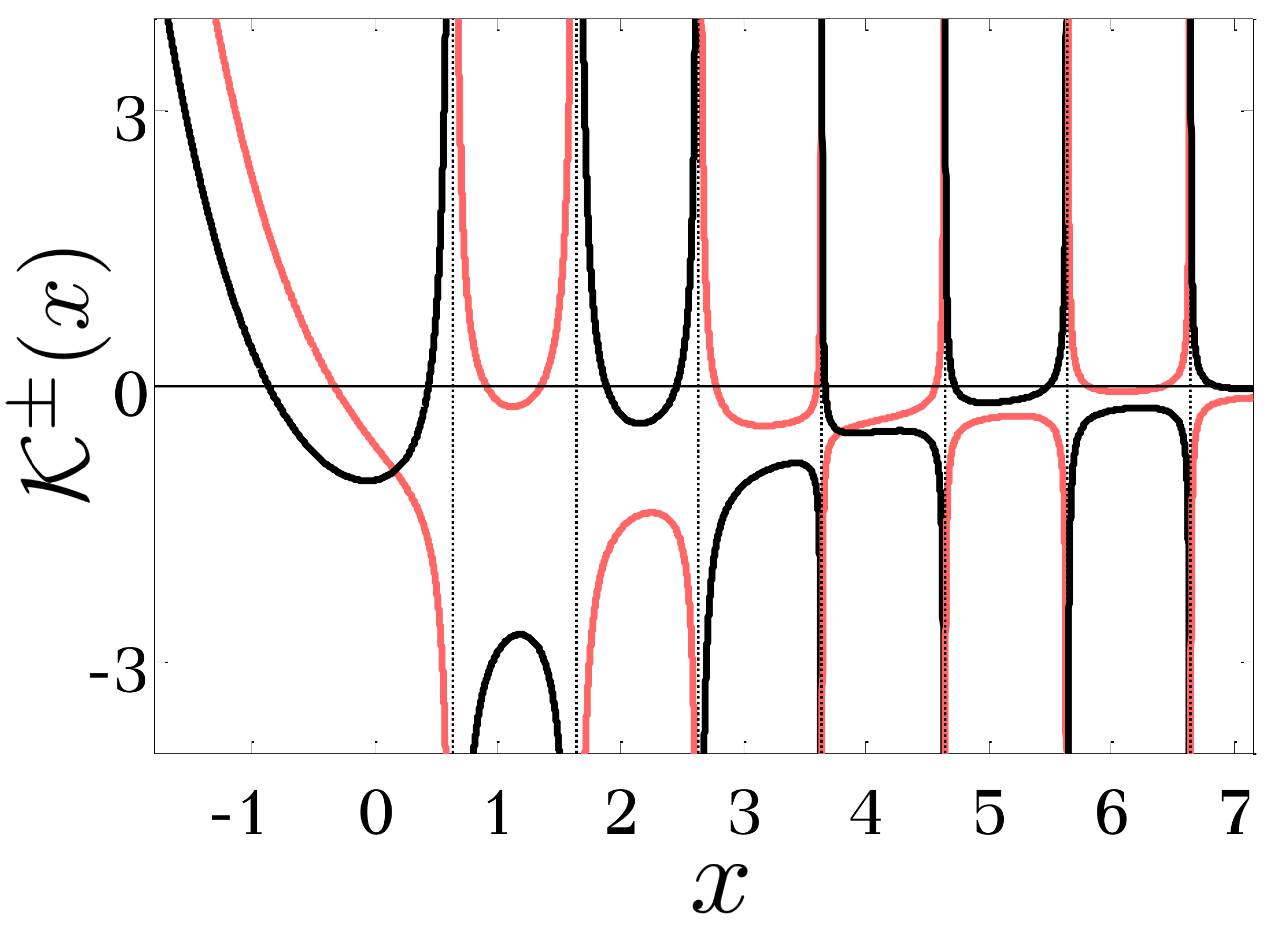}
\centering
\caption{$\mathcal{K}^{\pm}$ as a function of the continuous real parameter $x$ for $g=0.6$, $t=0.7$ and $\omega=1$. Black (positive parity) and red or gray (negative parity) curves correspond to the $\mathcal{K}^{+}$ and $\mathcal{K}^{-}$ in (\ref{D13}), respectively.}
\end{figure}

One can easily calculate the energy levels as a function of the electron-phonon interaction strength. Fig.~2 shows the energy levels as a function of $g$. From (\ref{D6}) it is easy to see that in the limit of very strong electron-phonon interaction, $g\gg 1$, where the electron hopping amplitude, $t$, can be ignored, the Hamiltonian is diagonal in the chosen basis and states with different parities coalesce to $\mathcal{E}^{k}=-\frac{g^2}{\omega}+k\omega$. The level crossings between two states with different parities is discussed in the next subsection (\MakeUppercase{\romannumeral 2}.C).
\begin{figure}[h]
\includegraphics[width=8cm]{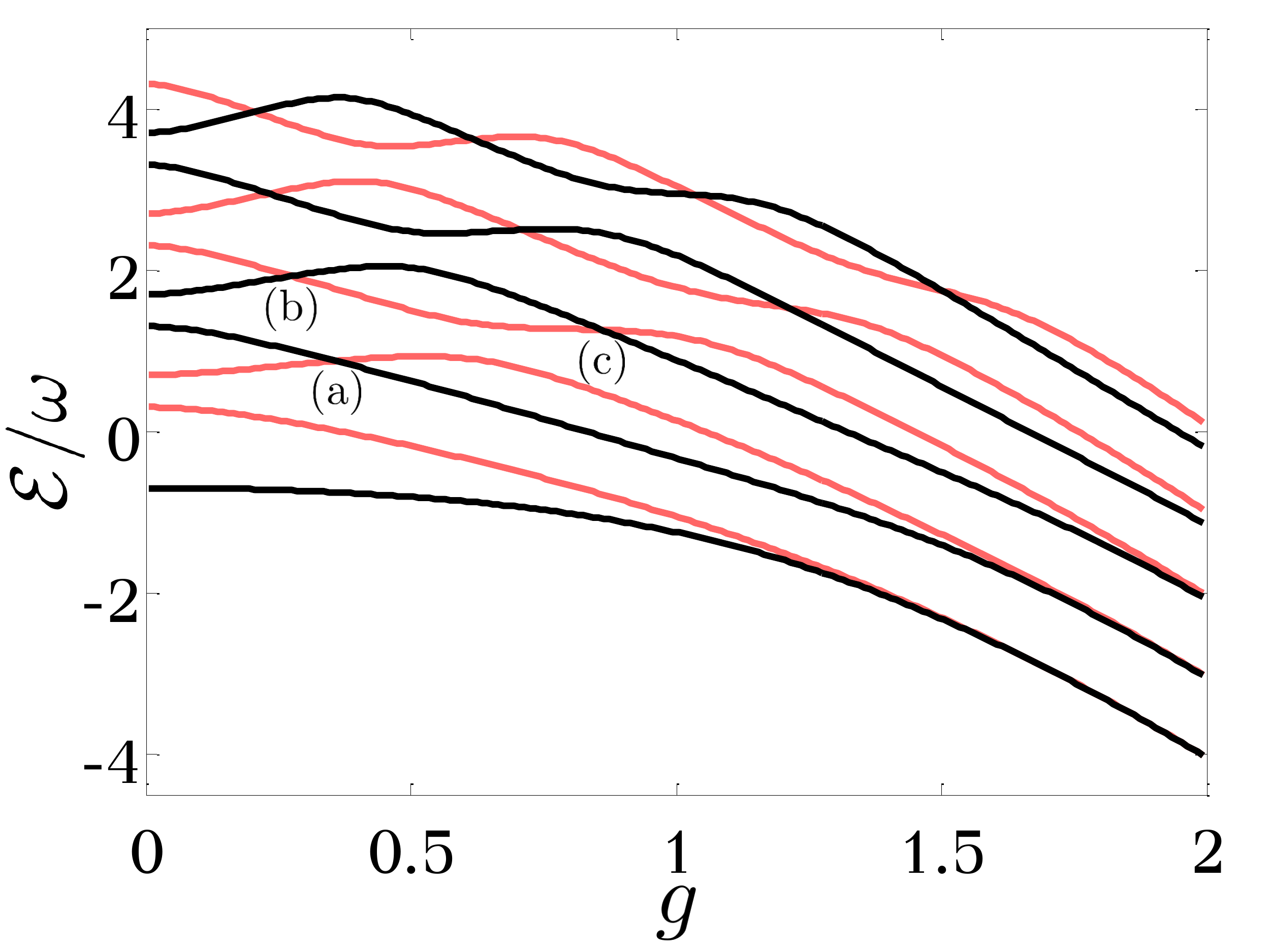}
\centering
\caption{Energy levels as a function of $g$ for $t=0.7$ and $\omega=1$. Black and red (gray) curves correspond to different parities.}
\end{figure}

In addition to the energy levels, other dynamical features, such as the lattice deformation \cite{Ranninger}, can be calculated effectively. The two time-independent correlation functions,
\begin{equation}
\mathcal{U}_{1}^{i}(g)=\langle c_{L}^{\dagger}c_{L}(b_{L}^{\dagger}+b_{L}) \rangle_{i},
\end{equation}
and
\begin{equation}
\mathcal{U}_{2}^{i}(g)=\langle c_{L}^{\dagger}c_{L}(b_{R}^{\dagger}+b_{R})\rangle_{i},
\end{equation}
where the subscript $i$ corresponds to the $i^{{\rm th}}$ eigenstate of the Hamiltonian (\ref{D1}), are measures of the on-site polaron-induced deformation and its extent over the secondary site, respectively. The explicit form of the functions are
\begin{align} \label{D14}
\mathcal{U}_{1}^{i}(g) &=\sqrt{2} \Big[ -\frac{g}{\omega}+\sum_{n=0}^{\infty} \sqrt{n+1}\, a_{L}^{i}(n)a_{L}^{i}(n+1)  \Big],   \nonumber \\
\mathcal{U}_{2}^{i}(g) &=-\sqrt{2} \sum_{n=0}^{\infty} \sqrt{n+1}\, a_{L}^{i}(n)a_{L}^{i}(n+1) .
\end{align}
Fig.~3 shows the correlation function $\mathcal{U}_{1}(g)$ for the ground and the first three excited states. The two highlighted points indicate the level crossing associated with point (a) shown in Fig.~2 where the second excited state becomes the third excited state and vice versa. As expected, on-site polaronic deformation strength, $\mathcal{U}_{1}(g)$, grows for all four states as $g$ increases. 
\begin{figure}[h]
\includegraphics[width=8cm]{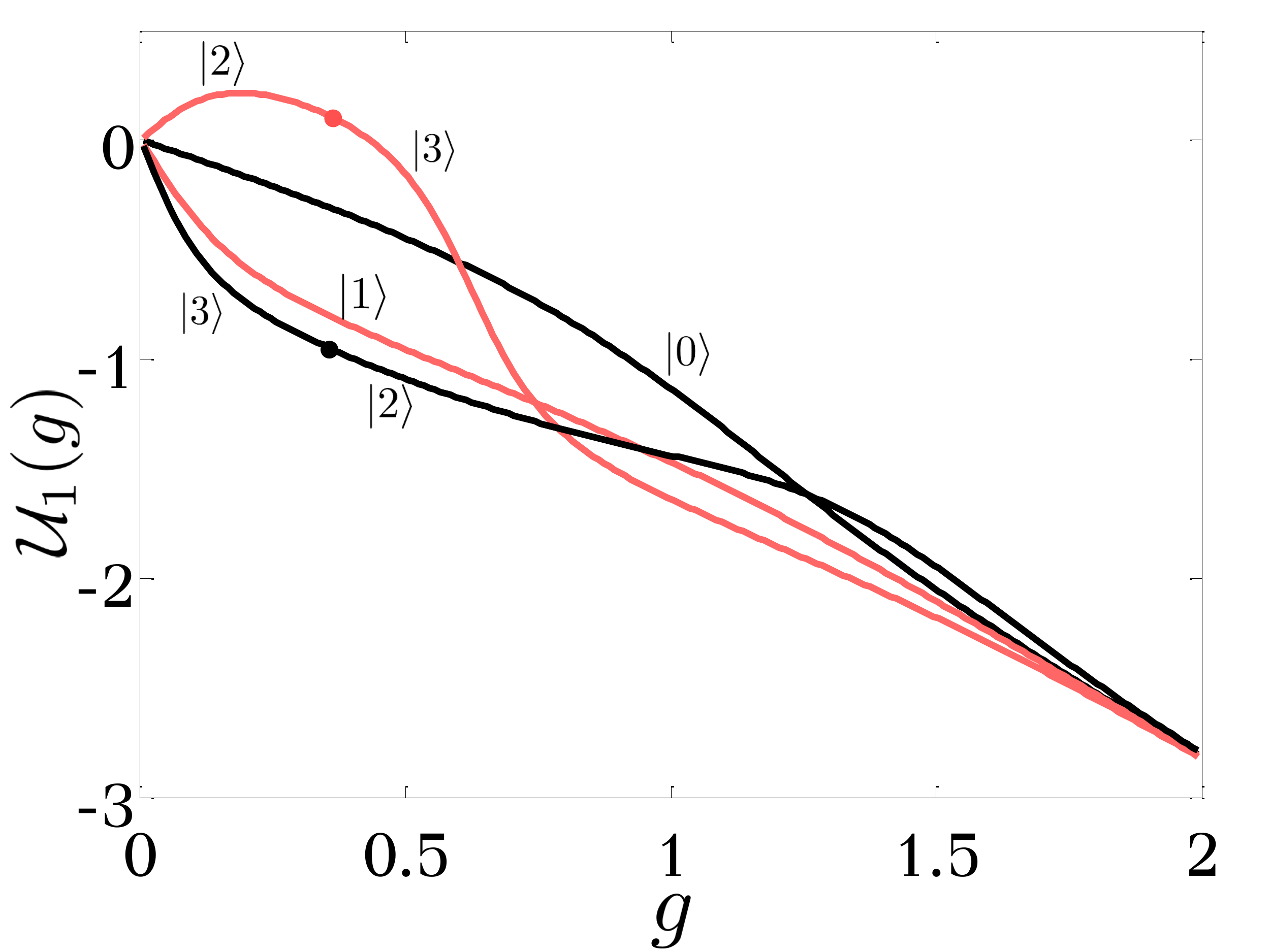}
\centering
\caption{Correlation function $\mathcal{U}_{1}(g)$ for the ground and the first three excited states ($t/ \omega=0.7$). The states in black and red (gray) have positive and negative parity, respectively.}
\end{figure}
Fig.~4 shows the correlation function $\mathcal{U}_{2}(g)$ for the first four states. Again, the points indicate the level crossing between states [point (a) in Fig.~2]. Fig.~4 illustrates the complexity of the lattice deformation range. For all states, as $g$ increases, the deformation extends to the neighboring site, i.e. the polaron size grows. However, for intermediate $g$, depending on the quantum number, different pictures emerge. For higher levels, it is possible to have no effect on the neighboring site for particular values of $g$. For strong $g$, however, regardless of the quantum state, the correlation vanishes. This indicates that the deformation is localized and hence the polaron becomes small.

\begin{figure}[h]
\includegraphics[width=8cm]{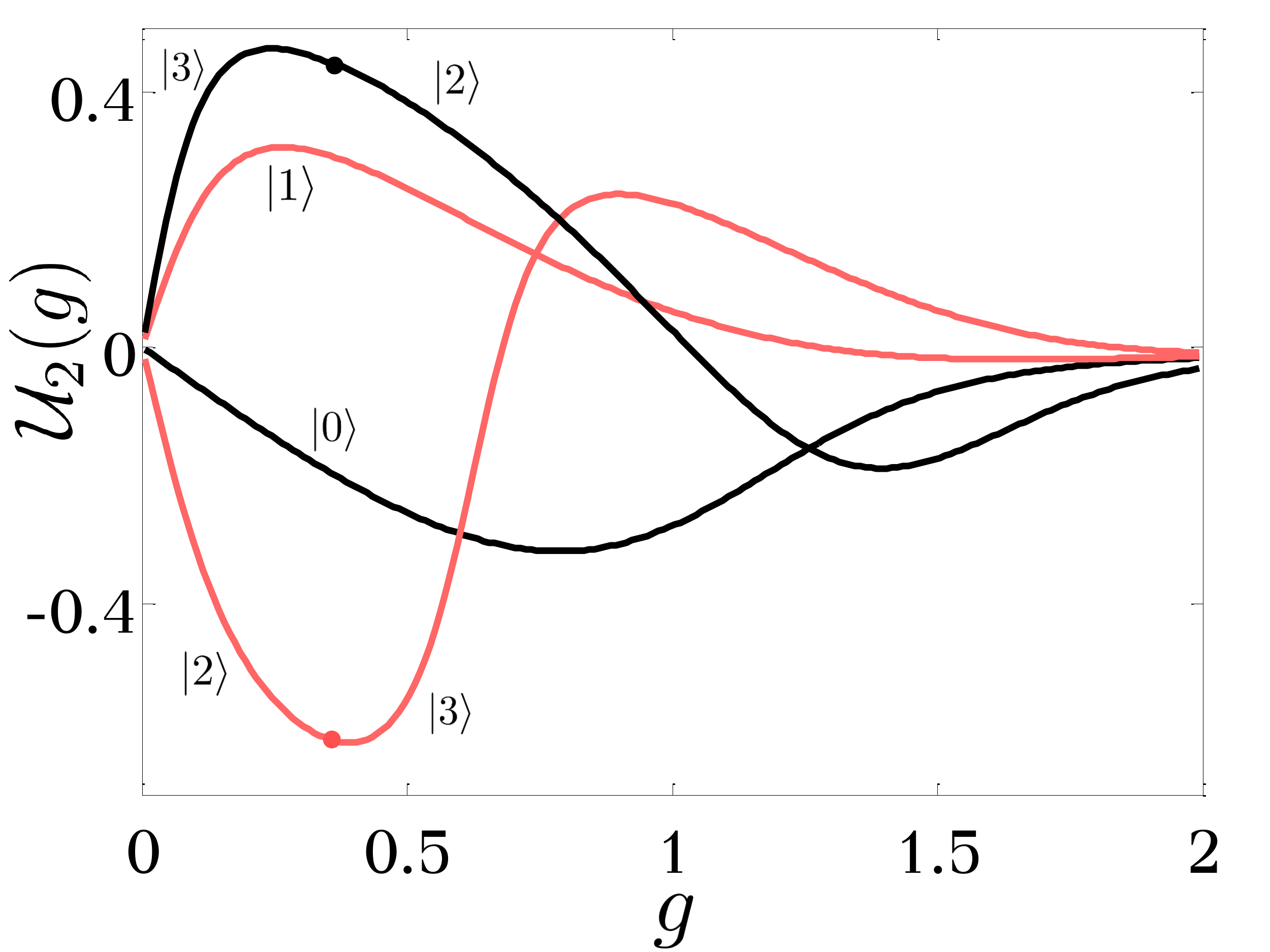}
\centering
\caption{Correlation function $\mathcal{U}_{2}(g)$ for the ground and the first three excited states ($t/ \omega=0.7$). The states in black and red (gray) have positive and negative parity, respectively.}
\end{figure}

\subsection{Level crossing}

For particular values of the model parameters, there exist analytical solutions. These isolated points are associated with Judd's exact solutions for the Jahn-Teller systems \cite{Judd}. Here, we obtain these isolated energy values and furthermore show that at these energies the symmetry of the eigenstates is broken and levels with differing parities cross. This is accomplished through the connection of the three-term recurrence relation with the confluent Heun differential equation (CHE). In fact, this connection was noted long ago in \cite{Reik} where it was conjectured that the wave functions of the Rabi Hamiltonian are given by a terminating series of generalized spheroidal functions. More recently in \cite{Zhong}, the eigenstates of the Rabi model were expressed in terms of the confluent Heun functions and these isolated solutions were discussed.

In order to find these solutions and the corresponding parameter values, the coefficients of the two three-term recurrence relations (\ref{D11}) and (\ref{C2}) are matched with one another. Here, (\ref{C2}) is the difference equation emerging out of the CHE (see appendix C). As a result of this matching, the parameters of the confluent Heun function, $H_{c}(\alpha, \beta, \gamma, \delta, q; x)$, are identified as
\begin{align}
\alpha&=\gamma-1=\delta=-\Big(\frac{g^{2}}{\omega^2}+\frac{\mathcal{E}}{\omega}\Big), \quad  \beta=-\Big(\frac{2g}{\omega}\Big)^{2}, \nonumber \\
q&=\Big(\frac{g^{2}}{\omega^2}+\frac{\mathcal{E}}{\omega}\Big)\Big(\frac{3g^{2}}{\omega^2}-\frac{\mathcal{E}}{\omega}\Big)+
\frac{t^{2}}{\omega^{2}}.
\end{align}
Given this, the recurrence relation (\ref{D11}) can be formally solved as
\begin{equation}
D(m)=\frac{1}{m!}\frac{d^{m}}{dx^{m}}H_{c}(\alpha, \beta, \gamma, \delta, q; x=0)
\end{equation}
Furthermore, we are equipped with a terminating condition for the recurrence relation (\ref{D11}) that puts an end to the otherwise infinite series of the wave function. The first condition in (\ref{C3}) provides the energy value of the terminated wave function, $\mathcal{E}=-\frac{g^{2}}{\omega}+N\omega$. The second condition in (\ref{C3}), however, results in an algebraic relation between the parameters $t$, $\omega$ and $g$. For instance, consider $N=1$, consequently $\mathcal{E}=-\frac{g^{2}}{\omega}+\omega$, and the parameters are related through $t^{2}+4g^{2}=\omega^{2}$. In Fig.~2 this is represented by point (a). Similarly, for $N=2$, the energy is $\mathcal{E}=-\frac{g^{2}}{\omega}+2\omega$ for the parameter values which satisfy $\Delta_{3}(q)=0$ in (\ref{C3}). The two points corresponding to this case are indicated as (b) and (c) in Fig. 2. At these special points levels of different symmetry cross and the energy eigenvalue is two-fold degenerate \cite{Braak,Zhong0}. The terminating condition results in the simultaneous termination of the amplitudes $a_{L}$ and $a_{R}$; at the exact crossing point, the symmetry is broken and the wave function has no definite parity.

\section{The four-site system} \label{secIII}

There has been an ongoing quest for an efficient method of solving the Holstein model with more than two sites \cite{Choudhury,Mezzacapo}. Usually, an LF-type transformation is performed before numerical calculations are carried out, which results in a more rapid convergence of the solutions. However, the complexity arises from dealing with exponential operators which leads to a dense matrix. Here, we show how our findings from the previous section lend themselves to effectively solving the four-site Holstein model. The symmetries inherited from the two-level system result in the formation of a sparse matrix, which is suitable for large-scale calculations. 

The Hamiltonian for a four-site system and a single electron is
\begin{equation} \label{D15}
H=t\sum_{i=1}^{4} c_{i}^{\dagger} c_{i+j}+ \omega \sum_{i=1}^{4} b_{i}^{\dagger} b_{i}+ g \sum_{i=1}^{4} c_{i}^{\dagger} c_{i}(b_{i}^{\dagger}+b_{i}) ,
\end{equation}
where $c_{i}^{\dagger} (b_{i}^{\dagger})$ and $c_{i}(b_{i})$ are fermionic (bosonic) creation and annihilation operators for site $i$, respectively. Here, $j$ runs over the nearest neighbor only. The chain is considered to be periodic. Next, with the help of the new phonon operators, we break the Hamiltonian into several two-site systems. Similar to \cite{Choudhury}, we define new phonon operators $p=(b_{1}+b_{2}+b_{3}+b_{4})/2$, $q=(b_{1}-b_{2}-b_{3}+b_{4})/2$, $s=(b_{1}+b_{2}-b_{3}-b_{4})/2$ and $r=(b_{1}-b_{2}+b_{3}-b_{4})/2$.
As a result, the transformed Hamiltonian is $H=H_{p}+H_{q,s}+H_{r}$, where
\begin{align} \label{D16}
H_{p} =& \omega p^{\dagger}p+\frac{g}{2}(p^{\dagger}+p)(c_{1}^{\dagger}c_{1}+c_{2}^{\dagger}c_{2}+c_{3}^{\dagger}c_{3}+c_{4}^{\dagger}c_{4}),   \nonumber \\
=& \omega p^{\dagger}p+\frac{g}{2}(p^{\dagger}+p),   \nonumber \\
H_{q,s} =& t(c_{1}^{\dagger}c_{2}+c_{2}^{\dagger}c_{1})+ \frac{\omega}{2} q^{\dagger}q+\frac{g}{2}(q^{\dagger}+q)(c_{1}^{\dagger}c_{1}-c_{2}^{\dagger}c_{2})  \nonumber \\
+ & t(c_{4}^{\dagger}c_{3}+c_{3}^{\dagger}c_{4})+ \frac{\omega}{2} q^{\dagger}q+\frac{g}{2}(q^{\dagger}+q)(c_{4}^{\dagger}c_{4}-c_{3}^{\dagger}c_{3})  \nonumber \\
+ & t(c_{1}^{\dagger}c_{4}+c_{4}^{\dagger}c_{1})+ \frac{\omega}{2} s^{\dagger}s+\frac{g}{2}(s^{\dagger}+s)(c_{1}^{\dagger}c_{1}-c_{4}^{\dagger}c_{4})  \nonumber \\
+ & t(c_{2}^{\dagger}c_{3}+c_{3}^{\dagger}c_{2})+ \frac{\omega}{2} s^{\dagger}s+\frac{g}{2}(s^{\dagger}+s)(c_{2}^{\dagger}c_{2}-c_{3}^{\dagger}c_{3}),  \nonumber \\
H_{r} =& \omega r^{\dagger}r+\frac{g}{2}(r^{\dagger}+r)(c_{1}^{\dagger}c_{1}-c_{2}^{\dagger}c_{2}+c_{3}^{\dagger}c_{3}-c_{4}^{\dagger}c_{4}).
\end{align}

Note that $H_{p}$ commutes with other parts of the Hamiltonian and hence can be diagonalized independently. The energy levels are $E_{n}=\omega n-g^{2}/(4\omega)$, where $n$ is a non-negative integer. The next piece, $H_{q,s}$, corresponds to four reduced two-level problems, similar to the Hamiltonian (\ref{D2}). Therefore, the exact solution of $H_{q,s}$ is developed using the previously constructed  solutions for the two-site system. The wave function is
\begin{align}
\psi^{i,j}= & c_{1}^{\dagger}\sum_{n=0}^{\infty}a_{L}^{i}(n) \phi_{q}^{+}(n) \sum_{n=0}^{\infty}a_{L}^{j}(n) \phi_{s}^{+}(n)  \nonumber \\
+ & c_{2}^{\dagger} \sum_{n=0}^{\infty}a_{R}^{i}(n) \phi_{q}^{-}(n) \sum_{n=0}^{\infty}a_{L}^{j}(n) \phi_{s}^{+}(n)  \nonumber \\
+ & c_{3}^{\dagger} \sum_{n=0}^{\infty}a_{R}^{i}(n) \phi_{q}^{-}(n) \sum_{n=0}^{\infty}a_{R}^{j}(n) \phi_{s}^{-}(n) \nonumber \\
+ & c_{4}^{\dagger} \sum_{n=0}^{\infty}a_{L}^{i}(n) \phi_{q}^{+}(n) \sum_{n=0}^{\infty}a_{R}^{j}(n) \phi_{s}^{-}(n) ,
\end{align}
where the amplitudes $a_{L}$ and $a_{R}$ are the same as in the previous section and $\phi_{q}^{\pm}$ and $\phi_{s}^{\pm}$ are eigenstates of shifted oscillators defined in accordance to Eqs.~(\ref{D4}) and (\ref{D5}). The associated eigenenergies are $E^{i,j}=\mathcal{E}^{i}(\omega,\frac{g}{2},t)+\mathcal{E}^{j}(\omega,\frac{g}{2},t)$ where the former and the latter are the $i$'th and $j$'th energy levels of a reduced two-level system with parameters $\omega$, $g/2$ and $t$.

Note that $H_{q,s}$ and $H_{r}$ do not commute and therefore $H_{q,s}+H_{r}$ has to be diagonalized by numerical means. The basis set is chosen to be $\{\psi^{i,j} \otimes \ket{n}_{r}\}$ where $\omega r^{\dagger}r\ket{n}_{r}=n\omega\ket{n}_{r}$. The matrix element of $H_{r}$ in this basis, $\bra{\psi^{i',j'} \otimes n'}H_{r} \ket{\psi^{i,j} \otimes n}$, is proportional to $\sum_{n=0}^{\infty}a_{L}^{i}(n)a_{L}^{i'}(n).\sum_{n=0}^{\infty}a_{L}^{j}(n)a_{L}^{j'}(n)$ which, due to the previous discussion, is non-zero only if $i$ and $i'$ and also $j$ and $j'$ belong to the same symmetry class. Therefore, the Hamiltonian matrix is extremely sparse. This leads to a computationally cost-efficient diagonalization of the matrix.
Fig.~5 shows the ground state energy for different adiabatic parameters. The results are in perfect agreement with existing literature \cite{Choudhury} and demonstrate the behavior of energy as a function of the electron-phonon interaction strength, $g$.
\begin{figure}[h]
\includegraphics[width=8cm]{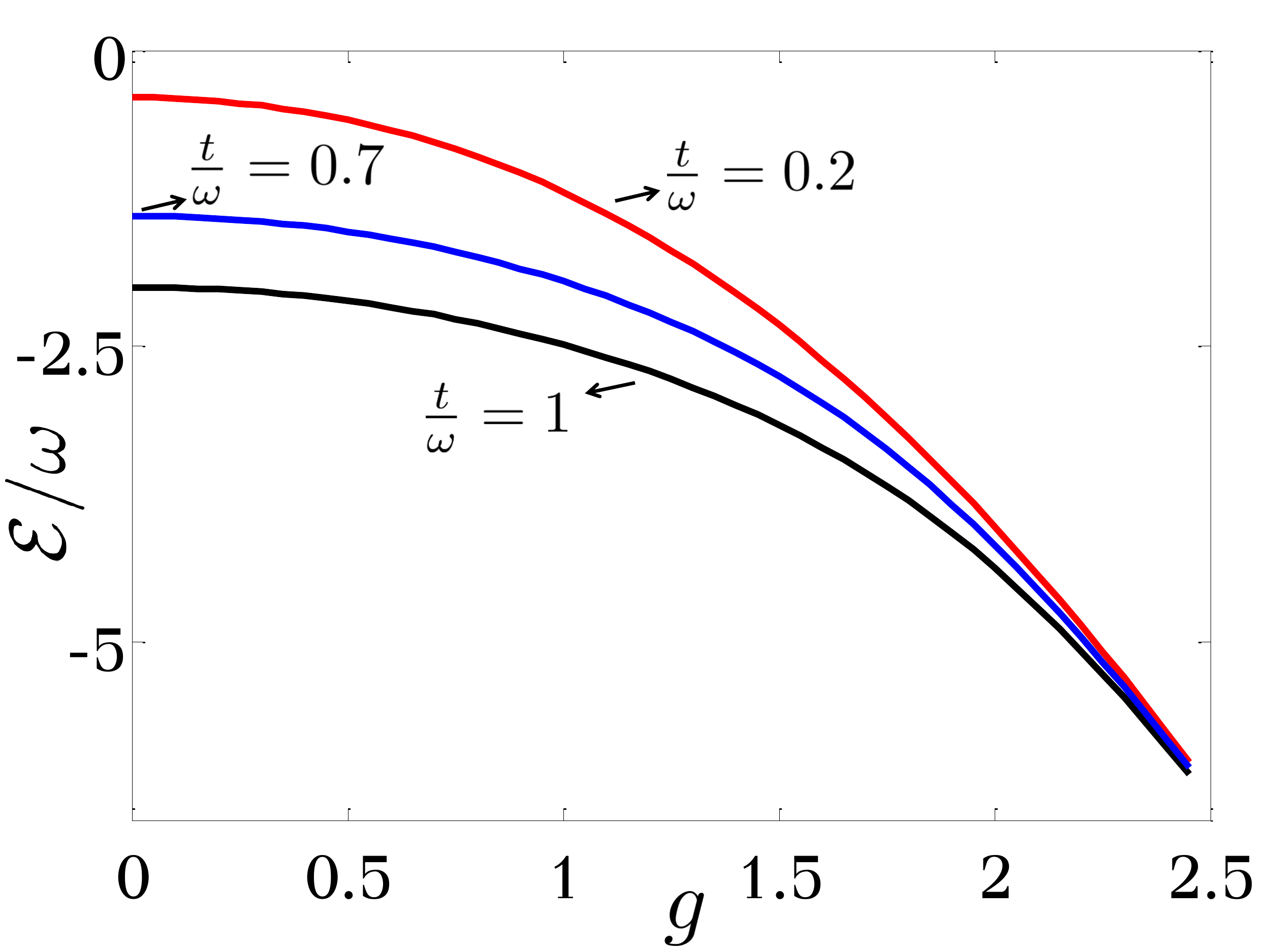}
\centering
\caption{Ground state of the four-site Holstein model (\ref{D15}) as a function of $g$ for various adiabatic parameters.}
\end{figure}

\section{Conclusion} \label{secIV}

The presented exact solution for the two-site Holstein polaron problem is not only useful for practical calculations but it is also suitable for further investigation of the symmetry, energy level crossings and polaronic characteristics of the Hamiltonian. A crucial finding was the three-term recurrence relation between the wave function amplitudes. This resulted in the efficient calculations of dynamical characteristics, such as energy levels and polaronic properties of the lattice for all values of the electron-phonon coupling. It was shown that the eigenstates fall into two groups with different symmetry (parity). The level crossing of these eigenstates was discussed in detail.

It was demonstrated through an example that the results obtained for a two-site system can form a convenient framework for studying systems with a greater number of sites. The Hamiltonian of a four-site Holstein model was broken up into several two-site pieces. This allowed for the extension of the obtained results to the case of the four-site system. The Hamiltonian matrix was found to be extremely sparse and easily suitable for numerical calculations.

It would be interesting to extend this framework to applications involving polaronic effects such as quantum transport. This is left for future studies.

\begin{acknowledgments}
A.T. thanks A. Stain for her support and helpful assistance. V.Z. acknowledges support from the NSF grant PHY-1404442. 
\end{acknowledgments}

\appendix

\section{Overlap functions $P^{+-}_{mn}$ and $P^{-+}_{mn}$}

In this appendix, we calculate the overlap functions $P^{+-}_{mn}$ and $P^{-+}_{mn}$ defined in Eqs.~(\ref{D7}) and (\ref{D8}), respectively. First, we recall the relations between the new phonon operators in Eqs.~(\ref{D4}) and (\ref{D5}), $\mathcal{B}_{+}=b+\lambda$ and $\mathcal{B}_{-}=b-\lambda$ (here $\lambda=g/\omega$); therefore, $\mathcal{B}_{+} =2\lambda+\mathcal{B}_{-}$. The ground states of the $B$-operators are related to the ground state of the $b$-operator ($b\ket{0_{b}}=0$) through the unitary displacement operator,
\begin{align} \label{A1}
\ket{\phi^{+}(n=0)}=e^{-\frac{1}{2}\lambda^{2}} e^{-\lambda b^{\dagger}} \ket{0_{b}} , \nonumber \\
\ket{\phi^{-}(n=0)}=e^{-\frac{1}{2}\lambda^{2}} e^{+\lambda b^{\dagger}} \ket{0_{b}} .
\end{align}
The first overlap function, $P^{+-}_{mn}=\bra{\phi^{+}(m)}\phi^{-}(n)\rangle$, equals
\begin{align} \label{A2}
P^{+-}_{mn} &=\frac{1}{\sqrt{m!}} \bra{\phi^{+}(0)} (\mathcal{B}_{+})^{m}\ket{\phi^{-}(n)} \nonumber \\
&= \frac{1}{\sqrt{m!}} \bra{\phi^{+}(0)} (2\lambda+\mathcal{B}_{-})^{m}\ket{\phi^{-}(n)} \nonumber \\
&= \frac{1}{\sqrt{m!}} \sum_{k=0}^{m} {m \choose k} (2\lambda)^{m-k} \sqrt{\frac{n!}{(n-k)!}} \nonumber\\
& \times \bra{\phi^{+}(0)}   \phi^{-}(n-k) \rangle .
\end{align}
In order to evaluate the last line in (\ref{A2}), we find using (\ref{A1}):
\begin{align} \label{A3}
\bra{\phi^{+}(0)}   \phi^{-}(n-k) \rangle &= \frac{e^{-\lambda^2}}{\sqrt{(n-k)!}} \nonumber\\
& \times \bra{0_{b}} e^{-\lambda b} (b^{\dagger}-\lambda)^{n-k} e^{\lambda b^{\dagger}} \ket{0_{b}} \nonumber \\
&= \frac{e^{-2\lambda^2}}{\sqrt{(n-k)!}} \ (-2\lambda)^{n-k} .
\end{align}
Substituting (\ref{A3}) back into (\ref{A2}), we obtain the final expression for the overlap function $P^{+-}_{mn}$:
\begin{align} \label{A4}
& P^{+-}_{mn} = \nonumber\\
& (-1)^{n} e^{-2\lambda^2} \sqrt{m!n!} \ (2\lambda)^{m+n} \sum_{k=0}^{min (m,n)} \ \frac{(-1)^{k} (2\lambda)^{-2k}}{k! (m-k)! (n-k)!} .
\end{align}
The second overlap $P^{-+}_{mn}$ can be evaluated in a similar manner,
\begin{align} \label{A5}
& P^{-+}_{mn} = \nonumber\\
& (-1)^{m} e^{-2\lambda^2} \sqrt{m!n!} \ (2\lambda)^{m+n} \sum_{k=0}^{min (m,n)} \ \frac{(-1)^{k} (2\lambda)^{-2k}}{k! (m-k)! (n-k)!} .
\end{align}

The overlap functions satisfy the following properties
\begin{align} \label{A6}
& P^{+-}_{mn} = P^{+-}_{nm}(-)^{m+n}, \quad P^{-+}_{mn} = P^{-+}_{nm}(-)^{m+n}, \nonumber \\
& P^{+-}_{mn} = P^{-+}_{mn}(-)^{m+n}. 
\end{align}

\section{Poisson-Charlier polynomials}

The sum $\sum \frac{(-1)^{k} (2\lambda)^{-2k}}{k! (m-k)! (n-k)!}$ in (\ref{A4}) and (\ref{A5}) can be written in terms of the Poisson-Charlier polynomials $c_{n}(\alpha;x)$ which form a family of orthogonal functions.
The Poisson-Charlier polynomials are defined as \cite{Bateman}
\begin{equation} \label{B1}
c_{n}(\alpha;x)=\sum_{k=0}^{n} (-1)^{k} {n \choose k}{\alpha \choose k}k!\ x^{-k}
\end{equation}
with $x>0$ and $\alpha=0, 1, 2, ...$. The polynomials are orthogonal with respect to the Poisson distribution,
\begin{equation} \label{B2}
\sum_{\alpha=0}^{\infty} \frac{x^{\alpha}}{\alpha!}c_{n}(\alpha;x)c_{n'}(\alpha;x)=x^{-n}e^{x}n!\delta_{nn'},
\end{equation}
and satisfy the recurrence relation \cite{NIST}
\begin{equation} \label{B3}
xc_{m+1}(n;x)+mc_{m-1}(n;x)=(m-n+x)c_{m}(n;x).
\end{equation}
Another important identity defines the generating function of these polynomials \cite{Bateman},
\begin{equation} \label{B4}
\sum_{n=0}^{\infty}c_{n}(\alpha;x)\frac{t^{n}}{n!}=\Big(1-\frac{t}{x}\Big)^{\alpha}e^{t},
\end{equation}
 where $|t|<x$. Using the definition in (\ref{B1}) one can write the overlap functions  (\ref{A4}) and (\ref{A5}) in terms of the polynomials:
\begin{align} \label{B5}
P^{+-}_{mn} &= (-1)^{n} e^{-2\lambda^2} \frac{(2\lambda)^{m+n} }{\sqrt{m!n!}} \ c_{n}\big(m,(2\lambda)^{2}\big) , \nonumber \\
P^{-+}_{mn} &= (-1)^{m}  e^{-2\lambda^2} \frac{(2\lambda)^{m+n}}{\sqrt{m!n!}} \  c_{n}\big(m,(2\lambda)^{2}\big) .
\end{align}

\section{Confluent Heun Equation}

The confluent Heun equation (CHE) is a second-order linear ordinary differential equation with two regular and one irregular singular points. The equation has multiple standard forms; here we adopt the non-symmetrical canonical form \cite{Ronveaux,Choun},
\begin{equation} \label{C1}
\frac{d^{2}y}{dx^{2}}+\Big(\beta+\frac{\gamma}{x}+\frac{\delta}{x-1}\Big)\frac{dy}{dx}+\frac{\alpha \beta x-q}{x(x-1)}y=0,
\end{equation}
where $x=0$ and $x=1$ are regular singular points and $x=\infty$ is an irregular singular point. The power series solution in the vicinity of the regular singular point $x=0$, $H_{c}(\alpha, \beta, \gamma, \delta, q; x)= \sum_{n=0}^{\infty}f_{n}x^{n}$,  yields a three-term recurrence relation
\begin{equation} \label{C2}
A_{n} f_{n+1}=B_{n} f_{n}+C_{n} f_{n-1} ,
\end{equation}
with the boundary conditions $f_{-1}=0$ and $f_{0}=1$. Here $A_{n}=(n+1)(n+\gamma)$, $B_{n}=n(n-\beta+\gamma+\delta-1)-q$ and $C_{n}=\beta(n+\alpha-1)$. The confluent Heun function, $H_{c}(\alpha, \beta, \gamma, \delta, q; x)$, reduces to a polynomial of order $N \geq 0$ if and only if the following two conditions hold simultaneously \cite{Fiziev}:
\begin{align} \label{C3}
\alpha+N &=0 , \\ \nonumber
\Delta_{N+1}(q) &=0 ,
\end{align}
where
\begin{widetext}
\begin{equation}
\Delta_{N+1}(q)=
\begin{vmatrix}
q-p_{1} & 1(\gamma) & 0 & \cdots & 0 & 0 & 0 \\
N\beta & q-p_{2}+1\beta & 2(1+\gamma) & \cdots & 0 & 0 & 0 \\
0 & (N-1)(\beta) & q-p_{3}+2\beta & \cdots & 0 & 0 & 0\\
\vdots & \vdots & \vdots & \ddots & \vdots  & \vdots  & \vdots   \\
0 & 0 & 0 & \cdots & q-p_{N-1}+(N-2)\beta & (N-1)(N+\gamma-2) & 0\\
0 & 0 & 0 & \cdots & 2\beta &  q-p_{N}+(N-1)\beta & N(N+\gamma-1)\\
0 & 0 & 0 & \cdots & 0 & 1\beta & q-p_{N+1}+N\beta\\ \notag
\end{vmatrix}
\end{equation}
\end{widetext}
and $p_{n}=(n-1)(n+\gamma+\delta-2)$.
The first condition in (\ref{C3}) guarantees the vanishing of $C_{N+1}$ whereas the second condition is equivalent to $f_{N+1}=0$. The two conditions together force the recurrence relation (\ref{C2}) to terminate.

\bibliography{Refs}

\end{document}